\documentclass{aa501}
\usepackage{graphicx}
\begin{document}
   \title{The Rees-Sciama effect and the primordial nucleosynthesis}

   \author{A. M\'esz\'aros}

   \institute{Astronomical Institute of the Charles University,
              V Hole\v{s}ovi\v{c}k\'ach 2, 180 00 Prague 8,
	      Czech Republic\\
              \email{meszaros@mbox.cesnet.cz}
              }
   \date{Received November 21, 2001; accepted  March 8, 2002}

   \abstract{
   It is known that, theoretically, the Rees-Sciama effect may cause
arbitrarily large additional redshifts in the cosmic
microwave background radiation due to transparent expanding 
voids having sizes comparable with the size of
horizon. Therefore, again theoretically, eventual huge voids existing 
immediately after the recombination may essentially change the 
predictions of the theory of big bang nucleosynthesis. If this eventuality
holds, then the dark matter can be dominantly baryonic and, simultaneously,
one can be in accordance with the predictions of primordial
nucleosynthesis theory. Studying this eventuality one arrives
at the result that the observed extreme isotropy of the 
cosmic microwave background radiation rejects 
the existence of any such huge voids, 
and hence this eventuality does not hold.
}

   \maketitle

\section{Introduction}

The Rees-Sciama ef\-fect (Rees \& Sciama (1968);
 for a detailed survey see M\'e\-sz\'a\-ros \& Mol\-n\'ar (1996),
Zaldarriaga et al. (1998), 
Sakai et al. (1999) and the references therein)
causes an additional non-Friedmann shift of the photons of
cosmic microwave background radiation
(hereafter CMBR) due to the changing gravitational field of
transparent expanding structures being between the
last scattering surface and us. These structures may be either
overdensities (cf. superclusters) or underdensities (cf. voids), and
also the additional shift may be both redshift or blueshift, respectively.

This effect may clearly be important in galaxy formation scenarios,
where the initial perturbations during the recombination play a cardinal role,
and these perturbations may be deduced from the observed anisotropies of CMBR.
The importance follows from the fact that,
due to the Rees-Sciama effect, the observed
anisotropies of CMBR need not entirely be caused by the inhomogeneities
of non-relativistic matter during the recombination, and the later
transparent structures may also cause anisotropies in CMBR 
(see M\'esz\'aros (1994) for more discussion of this question). 

In addition, surprisingly, this effect can have a connection 
to completely different cosmological topics. For example, 
M\'esz\'aros \& Moln\'ar (1996)
show an interesting connection between this effect
and the observations of Lauer \& Postman (1994), where the
peculiar velocity of the Earth and the maximum of the 
dipole anisotropy of CMBR were found to have different directions. 
M\'esz\'aros \& Moln\'ar (1996) show that - theoretically and surprisingly
- the dipole anisotropy of CMBR may also  
be caused by the Rees-Sciama effect. Nevertheless, further
observational data exclude this possibility. The importance of that
work was recognized even by the editors of 
Sky \& Telescope Magazine (\cite{st}).

A similar surprising connection is the subject of this article. It discusses
the connection between this effect and the primordial nucleosynthesis.
Interestingly, this connection may have an essential impact on the character 
of dark matter of Universe (at least, in principle), because
this dark matter can dominantly be baryonic. Because the character
of dark matter is one of the most important open questions of 
present-day Cosmology, it is doubtlessly true that the subject of this article
is highly topical.

The paper is organized as follows. Sect. 2 succinctly
summarizes the known facts
about the Rees-Sciama effect. In Sect. 3 the connection between this effect
and the primordial nucleosynthesis is formulated. 
Sect. 4 shows that observations reject this
connection. Finally, in Sect. 5, the results of paper are summarized.

\section{Survey of the Rees-Sciama effect}

In this Section, we summarize briefly the known
facts concerning the Rees-Sciama effect.

The majority of the several papers (see M\'esz\'aros \& Moln\'ar (1996)
and references therein) dealing with this effect
discusses the case of a single spherically symmetric overdense region.
There are only a few articles
that discuss the case of a single void (\cite{tv87}, \cite{me94},
\cite{mm96}).

Summarizing the main results of the topic one may conclude:
a. The effect is independent of wavelength, and therefore the
black-body spectrum remains, and only the corresponding temperature is
changed due to the crossing of photons across a transparent structure;
b. The void causes an additional redshift (i.e. the
temperature of CMBR is smaller than the Friedmann value);
c. The overdense region causes both additional redshift and blueshift,
and the profile of this change is calculable either analytically 
or numerically; d. Effects from several voids and superclusters
along the path of the photons of CMBR should simply be summed;
e. The order of the effect is $|\delta T/T| \sim (|\delta \rho|/\rho)
(y/d_{hor})^{\beta}$, where $T$ is the temparature of CMBR, $\delta T$ is its
change, $\beta \simeq (2.5 - 3.0)$, 
$\rho$ is the Friedmann density, $\delta \rho$ is its departure
from this value; $y$ is the size of the object causing the additional shift,
and $d_{hor}$ is the size of horizon. It is
essential to note that $\delta T$ and $T$ are the present-day observable 
values; the remaining quantities are understood for the time, when the object
is crossed by the CMBR photons. 
It is also necessary to note that $|\delta \rho|/\rho \sim 1$ may
also occur; i.e. the structures can be highly non-linear. The effect is
increasing roughly cubically by the size of object. To illustrate,
consider a spherical empty void with 
diameter $\sim 100 h^{-1}$ Mpc at $z \ll 1$; i.e. in the cosmological sense
very recent ($z$ is the redshift, and
$H = 100 h$ km/(s Mpc) is the Hubble parameter).
Then the effect is of order $|\delta T/T| \sim 10^{-5}$ (\cite{me94}).
It must also be noted that - at least theoretically
- this effect need not be so small. For example, Rees \& Sciama (1968)
in their original paper discuss a hypothetical case, when the effect is
of order $|\delta T/T| \simeq 2\times 10^{-3}$. It is even possible that
the effect can be of order unity; again in principle. For example,
assume that a void with $|\delta \rho|/\rho \simeq 1$ having a present-day
size $\simeq 300 $ Mpc existed already during the recombination epoch. Because
during the recombination epoch its physical size is $\sim 1000$ times
smaller, at that time $d_{hor}$ was at the same order. In this
hypothetical case this void would give an effect of order unity.

\section{A possible impact 
on the character of dark matter and on the primordial nucleosynthesis}

Surprisingly,
in principle, the Rees-Sciama effect may lead to a new 
argument supporting the baryonic character of dark matter
based on the theory of primordial 
nucleosynthesis. As far as it is known, this connection was
never mentioned yet. 

The key idea of this connection is the following.
About 90 \% of matter in the Universe is dark, and
recently is widely accepted that this dark matter is dominantly
non-baryonic (\cite{pee93}).
The key argument against the baryonic character of
dark matter is based on the theory of nucleosynthesis.
This argument follows from the coincidence between the prediction of this
theory and
the observed abundances of H, D, $^3$He, $^4$He and $^7$Li. A few years
ago it was believed that this coincidence
was fulfilled for the ratio $\eta = n_b/n_{\gamma}
= (1.6 \pm 0.1) \times 10^{-10}$, where $n_b$ is the number density of
baryonic matter, and $n_{\gamma}$ is the number density of photons
of CMBR (\cite{dar95}). Later it was suggested
(\cite{tu96}, \cite{cf96}, \cite{ste99}) 
that this small and relatively pre\-cise value was not
cor\-rect. For example, Steigman et al. (1999)
obtain a much bigger value for this ratio from the 
primordial nucleosynthesis; the value $\eta 
^{>}_{\sim} 6 \times 10^{-10}$ is favored, and even the value 
$\eta = 13 \times 10^{-10}$ is not excluded. On the other hand, a value
as small as $\eta \simeq 2 \times 10^{-10}$ is not yet excluded either.
Recently one has $n_{\gamma} = 410 cm^{-3}$, and
$n_b = 1.124 \times 10^{-5} \Omega_b h^2 cm^{-3}$,
where $\Omega_b$ is the
ratio of baryonic density to the critical one (\cite{pee93}; p.103).
Hence, one has $n_b/n_{\gamma} = 2.74 \times 10^{-8} \Omega_b h^2$.
To be in accord with the earlier values of the theory of nucleosynthesis
it must be that $\Omega_b h^2 = (5.8 \pm 0.4) \times 10^{-3}$; while
with the highest allowed value of Steigman et al. (1999) 
$\Omega_b h^2 \simeq 4 \times 10^{-2}$.
The newest studies (see Coc et al. (2001) and the references therein) give 
$\Omega_b h^2 = (1.5 \pm 0.3) \times 10^{-2}$, i.e. $\eta = 4 \times 10^{-10}$.
Different observations suggest that $\Omega_M
\simeq (0.1 - 0.4)$, where $\Omega_M$ is the ratio of total
density of non-relativistic matter to
the critical density (\cite{bf98}). (Note here that for the purpose of this
paper it is completely irrelevant, if the cosmological constant is
zero or non-zero (\cite{ri00}).) This means that one must have
$\Omega_M/\Omega_b \simeq (2.5 - 80) h^2$, i.e. for the 
allowed $h \simeq (0.6-0.75)$ (\cite{free01}) the ratio
$\Omega_M/\Omega_b$ must be $\simeq (1 - 50)$. The first value 
seems to be excluded; only in the case of the lowest allowed
$\Omega_M h^2$ case with the highest allowed value of $\eta$ can this
occur. Because
about 90 \% of matter is dark, the situation is straightforwardly solvable,
if one assumes that this dark matter is dominantly non-baryonic.

For the sake of completeness it must be added here that - beyond the argument
based on the primordial nucleosynthesis - there are two other
different and independent arguments for the non-baryonic character of
the dark matter. The first argument is based 
on the observed anisotropies of CMBR (see, cf., de Bernardis et al. (2001)
and references therein). The obtained value $\Omega_b h^2 = (0.022 \pm 0.004)$
clearly needs a dominantly non-baryonic dark matter.
In addition, if one assumes that the dark matter 
is dominated by baryonic dark matter then any galaxy formation theory
is in doubt.
(For example, Peebles (1993) in Chapt. 25 "Baryonic Dark Matter" considers
the adiabatic dark matter scenario as "only of historical interest" and
also the isocurvature dark matter scenario is taken as an "unattractive"
scenario.) The second argument
follows from the observations of Lyman-$\alpha$ forest at high redshifts
giving $\Omega_b h^2 = (0.035 \pm 0.015)$ (see Hui et al. 2001 and
references therein). However, the discussion of these 
arguments is not a subject of this article.

Consideration in primordial nucleosynthesis is based on the key assumption that
$\eta = n_b/n_{\gamma} =$ constant during the whole expansion
of Universe. In the Friedmann models both $n_b$ and $n_{\gamma}$ must not
depend on spatial coordinates; they depend only on time $t$. Because
the functional dependence on time should be the same for both densities -
specifically 
one should have $n_b \sim n_{\gamma} \sim a(t)^{-3}$ ("the comoving
densities of CMBR photons and baryons, respectively, are
constant"; $a(t)$ is the expansion function) - 
the ratio of two densities should be
constant (for more details see, e.g., Weinberg (1972)).
(Departures from $a^3 n_{\gamma} =$ constant may
exist but should be small. For example, departures of order $\sim 10^{-5}$
should surely exist (\cite{pee71}; p. 232). 
$a^3 n_b$ should be even more constant.)
Strictly, the theoretical prediction of
$n_b/n_{\gamma}$ from the theory of primordial nucleosynthesis
is a prediction for this ratio for the first
a few minutes of Universe, when the temperature of CMBR is $\simeq
(1-10)$ MeV (Dar 1995). 

In addition, it is automatically assumed that
the ratio $n_b/n_{\gamma}$ does not change in the later stages
of Universe. Trivially,
if $n_b/n_{\gamma} =$ constant was not true in early times, then
this standard assumption would be essentially modified.

Some attempts have been made not to fulfil this assumption. One
of that is the case when $n_b$ depends on spatial coordinates. These
attempts do not lead to essentially new conclusions (\cite{jf95}).

Contrary to these attempts, assume for the moment that $n_b/n_{\gamma}$ is
changing drastically with time, but that spatial dependence is negligible.
If this
is the case, then the baryonic character of dark matter may easily be saved
under some specific conditions. Simply, one has to assume
that during the era of primordial nucleosynthesis $n_b/n_{\gamma} \simeq
10^{-10}$, but recently this ratio is $\simeq (2.5 - 80) h^{-2}$ times
bigger. This skip may occur - at least in principle - in three
ways: either by increasing the comoving number density
of baryons, by decreasing the comoving number density of CMBR
photons, or by the combination of both.
From the physical point of view only the case when the comoving 
number of photons is decreasing seems to be allowable. 

The key idea of this paper is to remark that such a decrease of the
number of photons may occur.
This decrease is in principle possible due to a global
Rees-Sciama effect of order unity caused by voids having the sizes
of horizon. This may be seen as follows.

If CMBR is a blackbody radiation (this is assumed everywhere in this
article), then $n_{\gamma}(t) \propto a^{-3}(t)$.
This relation is standardly assumed
to be fulfilled with high accuracy during the whole
epoch of expansion after the annihilation of electron-positron pairs
(\cite{pee71}, \cite{we72}).
Assume for the moment that this relation is
fulfilled only before recombination, but that does not hold
during a short period after recombination. Everything else is identical
to the case of the standard Friedmann model. In the standard picture
recombination occurs at $a_o/a(t) \simeq 1000$ (\cite{pee71}),
where $a_o$ is the present value of expansion function. Assume here that 
recombination occured later: at $a_o/a(t) = 1000/q$.
The value $q$ should be chosen to have
$q \simeq (1 - 50^{1/3}) \simeq (1 - 3.7)$. Before 
recombination the value of $\eta = n_b/n_{\gamma}$ was in accordance
with the prediction of primordial nucleosynthesis.
Nevertheless, immediately after recombination there is a
skip in the temperature of CMBR; this temperature is immediately
decreasing $q$-times. As the simplest situation, it may be
assumed that this skip occurs instantaneously; i.e.
during this decrease of temperature
the change of $a(t)$ is unimportant. Hence, if
this skip of temperature occurs, it can easily be that
$\Omega_M = \Omega_b$ today.

All this gives the following idea. Assume that, shortly
after recombination, the Universe is
filled by sufficiently large voids. This means that immediately
after recombination large non-linear structures exist.
(It must be stated here that
this situation is not precisely identical to the model of Jedamczik
\& Fuller (1995). They assume non-linear structures already during 
nucleosynthesis - i.e. during the first a few minutes of Universe. Here
no non-linearities are needed before recombination. They are needed only
shortly after recombination.)
This si\-tuation may lead - shortly after recombination -
to an additional global redshift; compared with
the Friedmann value the temperature of CMBR should decrease $q$ times.
Knowing the results of the studies concerning the
Rees-Sciama effect, one may conclude that this decrease is
possible - at least, in principle.

Of course, several questions emerge for this eventuality; both
from the theoretical and the observational point of view. The discussion
of these questions is the subject of the next Section.

\section{Solution: No impact}

From the purely theoretical point of view, the eventuality of the decrease
of $T$ of CMBR $q$-times with respect to the Friedmann value seems to be
allowed.

Theoretically, two conditions are required to occur for this phenomenon.
First, to have this huge Rees-Sciama effect 
the existence of voids of sizes comparable with the horizon at 
$z \simeq 1000/q \simeq (300 -1000)$, when the recombination occurs, 
is required. Remarkably, this can happen, because in the baryon dominated
galaxy formation scenarios the non-linearities should exist even
during recombination (M\'esz\'aros (1997) and references therein). 
Second, recombination should
be "delayed" and should occur at $z \simeq (300 - 1000)$.
The "delayed" recombination is in principle again allowed. 
On the other hand, there are also strange artificial
requirements: First, why exactly $q$ times does this additional 
decrease occur, and not - say - ten or hundred times more? Second, 
how did these huge voids arise already during recombination?
Here one has to add that some objects may exist during recombination
(\cite{me91}, \cite{me97}, \cite{bac98}, \cite{sak99}), but
not huge voids comparable with the size of horizon. Third, if these huge 
voids existed at $z \simeq (300 - 1000)$, why are some
objects not seen already at these redshifts? Hence, purely
from the theoretical point of view, the huge global Rees-Sciama effect is
allowed, but its occurence is strange enough.

Even worse is the situation from the observational point of view. There are
at least three counter arguments here. First, voids
of the sizes of the horizon at $z \simeq (300 - 1000)$ should today have
sizes of hundreds of Mpc. No such objects are observed; the
observed voids have sizes of tens of Mpc. Second,
the sizes of these voids - leading
to the global additional redshift defined by $q$ - are highly artificial,
because they should have roughly the same sizes. Nothing like this is
observed. Third, even if this were the case, it would be practically excluded
that in any direction the same huge (of order unity) additional
redshift occurs. The observed variance of this effect should be
of order $\sim 10^{-5}$ or smaller, 
because this is the order of the obser\-ved varian\-ce of
$|\delta T|/T$ (\cite{pee93}, \cite{zal98}, \cite{db01}). 
In fact, the ques\-tion of the varian\-ce of the Rees-Sciama ef\-fect - 
caused by se\-veral ob\-jects - was already discussed and solved in
de\-tail (\cite{tv87}, \cite{sak99}). 
The\-se ar\-tic\-les arrive at the result that the variance
and the size of the Rees-Sciama 
effect should ha\-ve the same or\-der. Therefore, voids with
sizes comparable to the horizon at $a_o/a(t) = 1000/q$ should give 
angular scales at $\simeq q$ degrees (this is the corresponding angular scale
for the horizon at these redshifts (\cite{pee93})) 
a variance of order unity in the temperature of the
CMBR. This is obviously not observed.
This is the key argument against the huge global Rees-Sciama effect.

It seems that any further speculation about the voids having
the sizes comparable with horizon scale is hopeless.

\section{Conclusions}

The purpose of this paper was to discuss the potential impact of the
Rees-Sciama effect on the character of dark matter. The first result
of this paper is the surprising fact that - in principle - once
a global Rees-Sciama effect occurs after the recombination, then the
dark matter may be dominantly baryonic. The second result is even more
surprising: from the purely theoretical point of view, under highly
artificial but physically allowed
assumptions, the global Rees-Sciama effect can really
occur. Nevertheless, the third result seems 
to be unambiguous: there is a clear contradiction
with the observations, because the extreme observed isotropy of CMBR on
the order $\sim 10^{-5}$ excludes the occurence of a global huge
Rees-Sciama effect.

One has still to add that - due to the fully negative
conclusion - it might appear that the
purposes of this paper were useless. The author argues
that this is not the case for three reasons. First,
any, in principle, allowed
connections between two different topics should always be discussed.
(In fact, the situation is similar to the case when the Rees-Sciama effect
and the Lauer-Postman's observations were discussed together. There
negative results were also obtained, but the importance of that
discussion was obvious.) Second, from the purely theoretical point
of view, the global Rees-Sciama effect is allowed; under artificially
strange conditions, but it is allowed. This conclusion alone is remarkable
and fully new. Third, it is not excluded that in the future the
si\-tuation will change due to new observational effects. For example, if 
reionization (\cite{bal98}, \cite{wel99}) 
were confirmed at - say - $z\simeq (5-20)$, then
the observed extreme isotropy of CMBR would reflect the situation
after reionization, and not after recombination; then the key argument
against the huge global Rees-Sciama effect would be overcame.

\begin{acknowledgements} 
The useful remarks of Dr. F. Vrba and of 
the anonymous referee are kindly acknowledged.
This research was supported by Research Grant J13/98: 113200004. 
\end{acknowledgements}

\end{document}